\documentstyle[aps,prl,preprint]{revtex}

\begin{document}

\title{\bf Static avalanches and Giant stress fluctuations in Silos}

\author{Philippe Claudin, Jean-Philippe Bouchaud}

\address{ Service
de Physique de l'Etat Condens\'e, CEA-Saclay, Orme des Merisiers, 
91 191 Gif
s/ Yvette CEDEX, France }

\date{\today}

\maketitle

\begin{abstract} We propose a simple model for arch formation in silos. We show that small pertubations (such as the thermal expansion of the beads) may lead to giant stress fluctuations on the bottom plate of the silo. The relative amplitude $\Delta$ of these fluctuations are found to be power-law distributed, as $\Delta^{-\tau}$, $\tau \simeq 1.0$. These fluctuations are related to large scale `static avalanches', which correspond to long-range redistributions of stress paths within the silo.
 \end{abstract}
\pacs{PACS numbers: 46.10+z, 05.40 +j, 83.70 Fn}

\vskip1pc
\narrowtext

Although granular matter is very familiar, it does display extremely
interesting and unexpected features. For example, the stress distribution
below a heap of sand shows a counter-intuitive {\it minimum} right below the
apex of the pile \cite{Smid}, reflecting the presence of `arches' (or line of
forces) within the pile \cite{Edwards,BCC,BCCW}. Another very common geometry is
that of the silo. In this case, it is known since Janssen in 1895 that the 
weight $\cal W$ supported by the bottom plate of a tall silo is only a small
fraction of the total weight of the grains contained in the silo -- most of
the weight is `absorbed' by the side walls \cite{Brown,Nedderman}. Very recent
experiments actually reveal an even more striking effect: the `apparent'
weight $\cal W$ depends very sensitively on temperature \cite{Clement}. More
precisely, a variation of temperature of a few degrees induce rather erratic
and abrupt changes of $\cal W$ of the order of a several \%, while relative change
in the size of the grains due to thermal expansion is only $10^{-5}-10^{-4}$ !
Furthermore, these changes are of {\it both signs} and hysteretic: the apparent weight $\cal W$ is not a
unique function of temperature. Another aspect of this phenomenon is the following: repeating the same experiment of filling a silo with precisely the 
same quantity of beads lead to an apparent weight $\cal W$ which fluctuates by
more than $20 \%$ \cite{Clement}.

It has been long recognized that stress propagation within granular media is
strongly inhomogeneous, with clear `stress paths' appearing in birefringence
experiments. This leads to large stress fluctuations which have been
recently investigated experimentally \cite{Liu}, numerically and theoretically
\cite{Liu,Copper}, through a simple `scalar' model of stress propagation,
where only the $zz$ (vertical) component of the stress tensor is considered.
Although oversimplified, this model leads to predictions in qualitative and
perhaps quantitative agreement with observations. From a theoretical point
of view, it can be shown \cite{usinprep} that the full tensorial stress
propagation model can indeed reduce, in some particular limits, to the scalar
case considered by Liu et al. \cite{Liu,Copper}. 

The basic idea to explain the `giant' stress fluctuations induced by
temperature changes (or any other small perturbation) is that the network of `stress paths' is extremely sensitive to small perturbations. A stress path reaching a side wall (which
can be seen as a partly absorbing boundary for the stress) might change its
direction and rather reach the bottom plate, leading to a sudden
upward `jump' in $\cal W$. Conversely, the apparent weight can jump {\it down}
 as a path reaching the bottom plate is transformed into one colliding with the
side walls. 

The model we use to describe more quantitatively this effect is an extension
of the one considered in \cite{Liu,Copper} to allow for the formation of
arches. For a two dimensional packing, each `grain' is
labelled by two integer $(i,n)$ giving its horizontal and vertical position. [Extension of the model to three dimensions is left for future work]. Each grain supports the weight of its two upstairs neighbours, and shares its
own load randomly between its two downstairs neighbours. The corresponding
scalar equation for weight propagation is thus: 
\begin{eqnarray} \nonumber 
W(i,n) &=& q_+(i-1,n-1)W(i-1,n-1) \\
&+& q_-(i+1,n-1)W(i+1,n-1) + w \label{1}
\end{eqnarray}
where $w$ is the weight of the grains, assumed to be constant. $q_\pm(i,n)$
is the fraction of the weight transmitted to the the grain $(i\pm 1,n+1)$,
and is a random variable between zero and one, subject to the mass
conservation constraint $q_+(i,n)+q_-(i,n)=1$. The walls are at $i=\pm L$, and are simply defined by the fact that $q\pm(\pm L,n) \equiv 0$, {\it and} 
$q_\mp(\pm L,n)$ randomly chosen between zero and one; a fraction $1-q_\mp(\pm L,n)$ of the mass is thus `absorbed' by the wall. 

As shown by Liu et al. \cite{Liu}, when
$q_+$ is uniformly distributed, the resulting asymptotic weight distribution
is given by $P(x=\frac{2{W}}{\langle W \rangle})= x \exp -x$, where $\langle
W \rangle$ is the average weight. When $q_+=0$ or $1$, the distribution
becomes much broader, decaying only as a power-law $W^{-\frac{4}{3}}$ \cite{Copper}. 

We want to include the possibility that when the local shear on a given
grain is too strong, this grain can `lift off' from one of its downstairs
neighbours and therefore only transmit its load to the neighbour which is
in the direction of the shear. Since we work with a scalar model where
shear is a priori absent, we assume that the transmitted shear stress from
$(i,n)$ to $(i \pm 1,n+1)$ is also proportional to $q_\pm(i,n) W(i,n)$, and
thus postulate that if 
\begin{eqnarray}\nonumber
& &q_+(i-1,n-1)W(i-1,n-1) \\
&-&q_-(i+1,n-1)W(i+1,n-1) > {\cal R}_c W(i,n)
\end{eqnarray}
(where ${\cal R}_c$ is a certain threshold), then $q_-(i,n)=0=1-q_+(i,n)$,
meaning that the link to the left of $(i,n)$ is removed. In a symmetric
fashion, if
\begin{eqnarray}\nonumber
& &q_-(i+1,n-1)W(i+1,n-1)\\
&-&q_+(i-1,n-1)W(i-1,n-1) > {\cal R}_c W(i,n)
\end{eqnarray}
then $q_+(i,n)$ is set to zero. Note that for ${\cal R}_c =1$, the model of
Liu et al. is exactly recovered (all the links are present). Conversely, for ${\cal R}_c=0$, the $q_\pm$ can only take values $0$ or $1$.

This rule is interesting because it potentially leads to `static' avalanches
in the following sense: if ${\cal R}_c$ is slightly decreased, a link
somewhere is removed -- say $q_-(i,n)$. Then it is quite probable that the
threshold will also be exceeded on site $(i+1,n+1)$, thereby inducing 
$q_-(i+1,n+1)$ to vanish, and so on. In other words, very long arches can
suddenly appear or disappear when ${\cal R}_c$ is changed (see Fig. 3
below). 

Physically, ${\cal R}_c$ should grow with the density $\phi=n a^d$ of the
packing ($d$ is the dimension of space), where $n$ is the number of grains per
unit volume and $a$ the size of the grains. The reason is that obviously,
contacts are easier to removed in a loose packing. It is thus reasonnable to
assume that ${\cal R}_c$ will grow with temperature, with 
\begin{equation}
\frac{\delta {\cal R}_c}{{\cal R}_c} \simeq \frac{\delta a}{a} 
\end{equation}
${\cal R}_c$ presumably also depends on the friction coefficient between the
grains, and is expected to be reach ${\cal R}_c=1$ in the limit of zero friction.

The total weight on the bottom plate of the silo is obviously defined as:
\begin{equation}
{\cal W}= \sum_{i=-L}^L W(i,H)
\end{equation}
where $H$ is the height of the silo. Fig 1 shows ${\cal W}$ as a function
of the threshold ${\cal R}_c$ for a given sample (i.e. a given set of
$q_\pm$ uniformly chosen between zero and one \cite{RQU}) and averaged over many samples (thick line), for a silo of size $D=2L+1=61$, $H=610$. 
${\cal W}$ is a constant for small ${\cal R}_c$, and then rapidly grows with
${\cal R}_c$, but reveals enormous fluctuations around the
average trend. More precisely, we find that for a small change 
$\delta{\cal R}_c$ of the threshold, then with probability $\rho \delta{\cal R}_c \ll 1$ there is a `static avalanche' in the sense that $\cal W$ changes by a {\it finite amount}, which is independent of $\delta{\cal R}_c$, and with probability $1-\rho \delta{\cal R}_c$, $\cal W$ does not vary at all. This picture is very similar to the `shocks' in Burgers' turbulence \cite{Burgers}.
Thus the full statistics of the ${\cal W}({\cal R}_c)$ curve can be decomposed into two aspects: the {\it density} of avalanches, and the {\it amplitude} of these avalanches.

$\bullet$ The density $\rho({\cal R}_c)$ of `avalanches' or `shocks' was obtained by plotting the probability that no shocks occur in an interval of length $\delta{\cal R}_c$, which is very well approximated by $\exp -[\rho \delta{\cal R}_c]$ (independent shocks). For $q$ uniformly distributed between
zero and one, we found that $\rho({\cal R}_c)$ is of the form $HL f({\cal R}_c)$, where $f$ is a function of order $1$ with increases rather rapidly with ${\cal R}_c$. Typically, for $H=200$ grain size and an aspect ratio of $1$, the density of shocks is $\sim 10000$. 

$\bullet$  The {\it relative} value of the jump of ${\cal W}$ when an `avalanche' occurs, will be denoted as $\Delta$. Interestingly, for an aspect ratio of $1$, but independently of ${\cal R}_c$ (and thus of the density $\rho$), $\Delta$ is distributed {\it according to a power-law} ${\cal P}(\Delta) \propto |\Delta|^{-\tau}$, with $\tau = 1.02 \pm 0.03$ (see Fig. 2).
On the largest sample ($H=D=201$), this power law is observed on nearly five decades, cut-off for $\Delta$ smaller than a certain $\Delta^*$ (which decreases with $H$), and cut-off above $\Delta=1$, which corresponds to relative change of the weight of $100 \%$.

From a practical point of view, it means that for $H=D=201$ (in grain size units), a small change of $\delta{\cal R}_c$ as low as $10^{-5}$ will, with probability $\rho \delta{\cal R}_c \simeq 10^{-1}$ induce a relative change $\Delta$ of
 the apparent weight $\Delta$, distributed as $\Delta^{-\tau}$ up to $\Delta \simeq 1$ ! Since $\tau$ is small (close to $1$), huge relative changes of the weight between say $10^{-2} - 10^{-1}$ have a probability given by $\frac{1}{\log_{10} \Delta^*}$, which is of the order of $0.2$. Hence, small perturbations typically induce large responses.

Fig 3 shows the `stress paths' (i.e. grains on which the load a certain arbitrary value) for two  very close values thresholds such that ${\cal W}$ has changed dramatically. A small perturbation indeed induces a noticable rearrangment of the stress paths (which could perhaps be directly detected by acoustic emission \cite{Bark}). This phenomenology is actually very close to the one of
quantum conductance fluctuations in strongly disordered samples, where small
changes of the chemical potential or the magnetic field lead to a complete
change in the optimally conducting paths \cite{1dcond}. The log of the
conductance then also shows large variations. Another similar situation is that of `directed polymers' in random environments, where small changes of the local
(disordered) potential, or a small external force, can lead to a large-scale change in the ground state conformation \cite{M,P,HF,HHZ}. In the same family 
of problems, one should
actually cite spin-glasses, which have also been argued to be `chaotic', i.e.
very sensitive to temperature or disorder changes \cite{Bray,FH}.

An interesting consequence of our model is that in the region where small 
perturbations lead to substantial effects, there is a large fraction of
links which take a value $q_\pm = 0,1$. This is turn, according to the
arguments of \cite{Copper}, should lead to a power-law distribution $W^{-\frac{4}{3}}$ for the local stress on the bottom plate. Fig 4 shows that the stress distribution for ${\cal R}_c = 0.9$ can be fitted to $W^{-1.18}$. Note however that the arguments of \cite{Copper} assume no correlations between the $q_\pm=0,1$, which is not the case here since broken links appear precisely along arches. For ${\cal R}_c =1$, the distribution has exponential tails \cite{Copper}. We thus expect that these giant fluctuations tend to disappear when the density of the packing is increased, or when the friction between the grains decreases.

When the aspect ratio $\frac{H}{D}$ increases, the situation changes in the following way. When a static avalanche occurs `far' from the bottom plate, the resulting change of weight is screened out by the presence of the
(partly) absorbing walls. A simple argument allows one to understand that 
the avalanche size distribution ${\cal P}(\Delta)$ is now cut-off beyond
a certain $\Delta_{\max}$ which decreases as the aspect ratio increases. This is precisely what we observe numerically. Furthermore, one observes a progressive `localisation' of the stress on a unique path, which, for large aspect ratios and small ${\cal R}_c$, carries a finite fraction of ${\cal W}$. 

Finally, let us discuss the above phenomenon from a slightly different point of view. Instead of fixing the disorder (i.e. the $q_\pm$) and changing the threshold ${\cal R}_c$, one can fix ${\cal R}_c$ and change the disorder. This would correspond to repeating the same experiment of filling a silo, thereby
building different local arrangments of the grains. As expected by a kind of ergodicity argument, the resulting distribution of apparent rescaled weight $\frac{{\cal W}}{\langle{\cal W}\rangle}$ is quite broad. As soon as the aspect ratio exceeds $5$, this distribution is
very close to a gaussian and independent of $H$, with a relative root mean square of $\sim 20 - 30 \%$. This compares rather well with the experiments. In the same spirit, hysteretic effects can be accounted for by the fact that when a
contact is lost and then reformed when the temperature comes back to its
initial value, the new values of $q_\pm$ have no reason to be the same as
before, since the local environment has slightly changed.

We have also investigated this model in the heap (sandpile) geometry. When ${\cal R}_c=1$, the stress distribution at the bottom of the pile is maximum 
just below the apex of the pile. However, when ${\cal R}_c$ is less than a certain critical value, arches spanning the size of the system suddenly proliferate, and a {\it dip} at the center of the pile is created, as observed experimentally. The importance of long arches to understand the `dip' was stressed in \cite{Edwards,BCCW}. The present model could be a microscopic foundation of this idea.

As a conclusion, we have shown that a motivated extension of the model
proposed by Liu et al. to account for arch formation can explain the rather
spectacular giant stress fluctuations induced by small temperature changes.
The basic idea is that of long-range modification of stress paths induced by a
small perturbation, which can be seen as `static avalanches'. Actually, our
model is rather close to the large class of `SOC' (Self-Organized Critical)
models \cite{BTW,Grinstein}, initially proposed to describe true `dynamical'
avalanches in granular media. The irony would be that these scale invariant
avalanches should rather be looked for in static properties !

\vskip 0.5cm 
Acknowlegments. We want to thank M.E. Cates, E. Cl\'ement, J. Duran, J. Rajchenbach and J. Wittmer for many fruitful discussions.

\vskip 0.5cm
Figure Captions
\vskip 0.5cm
Fig 1: Evolution of the apparent weight ${\cal W}$ as a function of the threshold for a silo of size $D=61$, and $H=601$. The thick line is an average over 1000 samples, while the thin line is a particular realisation of the disorder.

Fig 2: Distribution of shock amplitudes $\Delta$, in log-log coordinates, for ${\cal R}_c=0.6$ and $D=L=201$ (thick line), and $D=L=21$ (dotted line). For the larger
sample, the distribution is very nearly given by $\Delta^{-1}$ between $10^{-6}$ and $10^{-1}$.

Fig 3: Chaoticity of stress paths. We show the large stress paths for two very close values of ${\cal R}_c$, separated by a shock, and $H=L=201$.  For the second value of ${\cal R}_c$, the paths have been translated by $x \to x+2$ for clarity. 
As one can see, some large scale features of the stress network have changed; the total weight carried by the bottom plate has changed ${\cal W}= 0.735$ to
${\cal W}= 0.845$ !

Fig 4: Distribution of the local stress $W_i$ on the bottom plate, for ${\cal R}_c=0.9$ and a sample of size $H=201$, $D=201$. The best fit gives an exponent $-1.18 \pm 0.04$ for the exponent. This value is not far from the value $-\frac{4}{3}$ expected from the analysis of Coppersmith et al. \cite{Copper}, valid when $q_\pm$ are independent and can only take the values $0$ or $1$.


\begin{thebibliography}{99}
\bibitem{RQU} More generally, $q$ is chosen uniformly in the interval 
$[\frac{1-\epsilon}{2}, \frac{1+\epsilon}{2}]$. ). As one could expect, all the 
results obtained below qualitatively hold independently of $\epsilon$. For example, the existence of a power-law regime, and the value of the exponent $\tau$ are independent on the strength of the disorder $\epsilon$.
\bibitem{Bark} This noise is similar to the Barhausen noise in magnets, which reflects the sudden change of the domain wall positions, much like the change of stress paths proposed here.
\end{thebibliography}
 
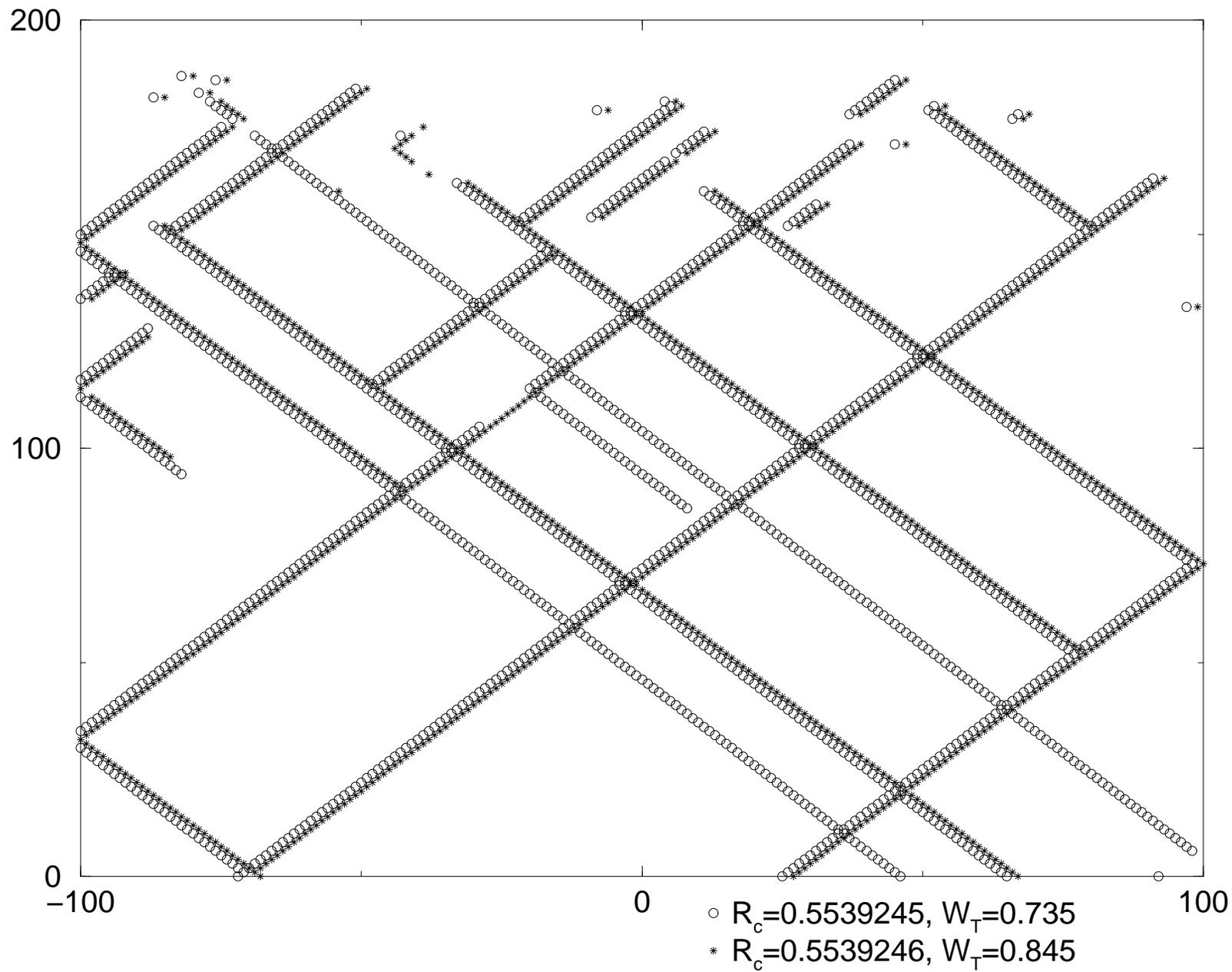
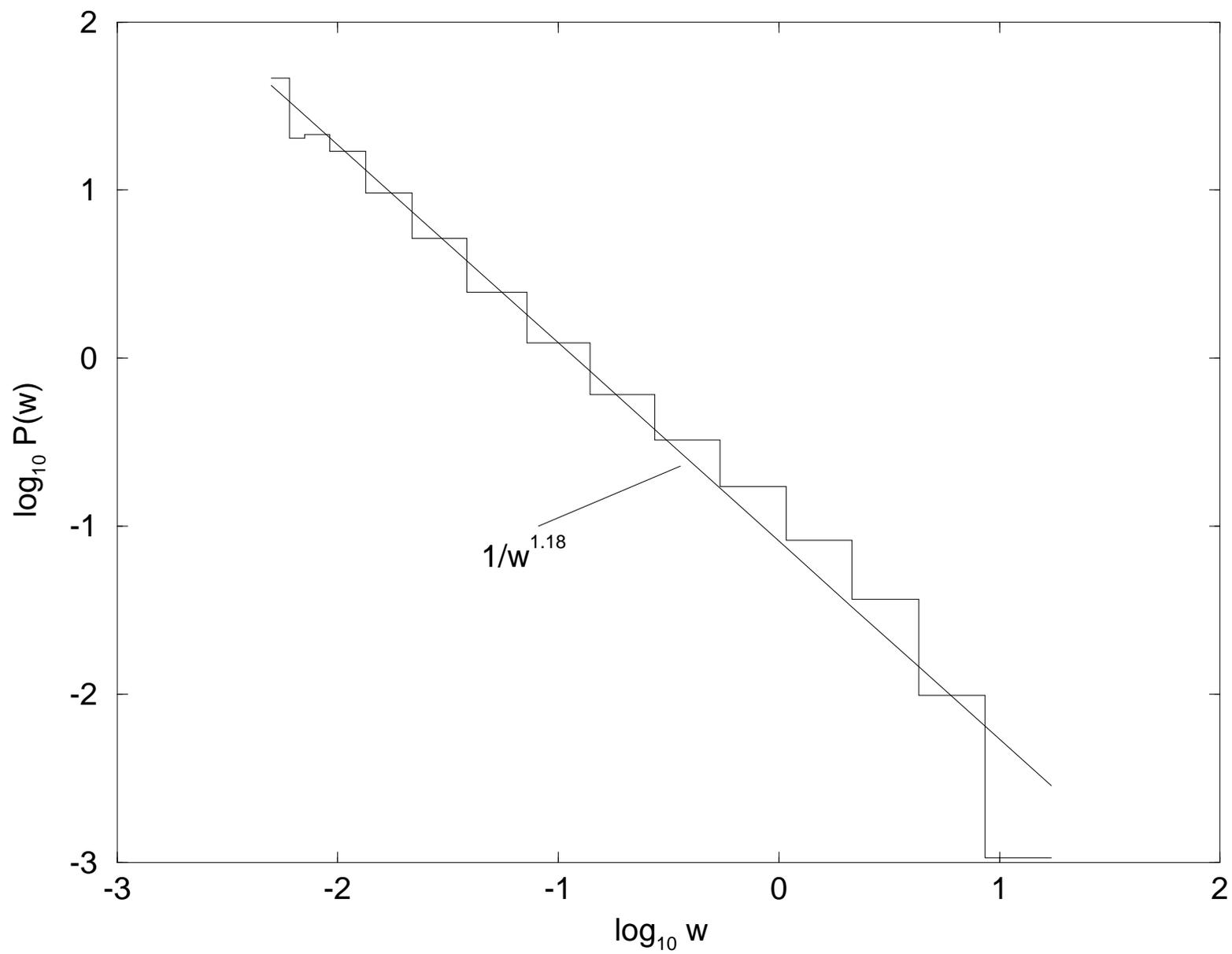
\end{document}